\documentclass[twocolumn]{aastex631}

\begin{document}
\preprint{APS/123-QED}
\title{Morphological Regimes of Rotating Moist Convection}
\author{Whitney T. Powers}
\affiliation{Department of Astrophysical and Planetary Sciences, University of Colorado Boulder, Boulder, CO 80309, USA}

\author{Adrian Fraser}
\affiliation{Department of Applied Mathematics, University of Colorado Boulder, Boulder, CO 80309, USA}

\author{Evan H. Anders}
\affiliation{Kavli Institute for Theoretical Physics, University of California, Santa Barbara, Santa Barbara, CA, 93106, USA}

\author{Jeffrey S. Oishi}
\affiliation{Department of Mechanical Engineering and Program in Integrated Applied
Mathematics, University of New Hampshire, Durham, NH 03824, USA}

\author{Benjamin P. Brown}
\affiliation{Department of Astrophysical and Planetary Sciences, University of Colorado Boulder, Boulder, CO 80309, USA}
%\date{August 2024}
\begin{abstract}
Moist convection is a physical process where the latent heat released by condensation acts as a buoyancy source that can enhance or even trigger an overturning convective instability. Since the saturation temperature often decreases with height, condensation releases latent heat preferentially in regions of upflow. Due to this inhomogeneous heat source, moist convection may be more sensitive to changes in flow morphology, such as those induced by rotation, than dry Rayleigh-B\'enard convection. In order to study the effects of rotation on flows driven by latent heat release, we present a suite of numerical simulations that solve the Rainy-B\'enard equations \citep{VPT2019}. We identify three morphological regimes: a cellular regime and a plume regime broadly analogous to those found in rotating Rayleigh B\'enard convection, and a novel funnel regime that lacks a clear analog within the regimes exhibited by dry convection. We measure energy fluxes through the system and report rotational scalings of the Reynolds and moist Nusselt numbers. We find that moist static energy transport, as measured by a moist Nusselt number, is significantly enhanced in the funnel regime without a corresponding enhancement in Reynolds number, indicating that this funnel regime produces structures with more favorable correlations between the temperature and vertical velocity.
\end{abstract}

%\maketitle

\section{Introduction}

Condensation releases latent heat which increases the buoyancy of a fluid parcel. Under certain conditions, the latent heat release is sufficient to drive convection in the absence of an unstable potential temperature gradient. The driving force behind this moist convective instability can be highly localized as condensation only occurs in those portions of the fluid which rise into regions where the saturation level is reduced. Thus, unlike dry Rayleigh-B\'enard convection, moist convection lacks up-down symmetry and may be more sensitive  to changes in flow morphology, such as those induced by rotation.

One such region where moist convection occurs and may be influenced by rotation is Jupiter's weather layer. This region is a thin region in the outer atmosphere with prominent ammonia and water clouds, and is where the planet's characteristic banded structure and stable geometrically arranged vortexes at the poles are observed \citep{Adriani2018Natur.555..216A}. Moist convection occurs in this region and has been linked to an upscale energy cascade in Jupiter's polar regions \citep{Siegelman2022NatPh..18..357S}. Additionally, moist convection may contribute to jet formation in the mid-latitudes \citep{Ingersoll2000Natur.403..630I, LIAN2010373} and the dearth of moist-convective activity near the equator may be responsible for the overabundance of ammonia in this region \citep{Guillot2020JGRE..12506404G}.

Upflows driven by moist convection can cause storms, and the location of these storms can then be mapped by observations of lightning strikes. Observations from the Galileo and Juno missions found that lightning strikes, and thus moist convective motions, are more common at latitudes above 40 degrees and are more common in the northern hemisphere \citep{Brown2018Natur.558...87B, LITTLE1999306}. This hemispherical asymmetry is present in both data sets, and is present throughout the Juno observations, suggesting that this is not an artifact from an anomalously strong storm. Furthermore, these studies found that moist convection is more prevalent in regions of cyclonic shear. The mechanisms behind the global distribution of moist convection are not fully known, and study of the fundamental properties of latent heat-driven convection is an important step towards understanding the dynamics of Jupiter's atmosphere.

Jupiter's dynamics are strongly influenced by rotation. Jupiter has a rotation period of approximately eight hours, and estimates place the Rossby number, the ratio of inertial and  Coriolis forces, in the interior at $10^{-5}$ to $10^{-4}$ and in the weather layer at $0.4$ \citep{Guillot_FransTextbook_2004jpsm.book...35G}. Since the Rossby number of the weather layer is less than one, the effects of rotation are significant and must be included in models that aim to study this region. 

 Dry, rotating, Rayleigh-B\'enard convection has been studied in great detail. As the reduced Rayleigh number, $\mathcal{R}$ (a dimensionless measure of buoyant driving relative to the stabilizing effects of viscosity, diffusion, and rotation) is increased, the system transitions through several morphological regimes, each with their own characteristic structures, force balances, and energetics. At reduced Rayleigh numbers just above onset, rotating Rayleigh-B\'enard convection exhibits a cellular regime characterized by laminar upflow and downflow columns packed in a regular hexagonal pattern \citep{Chandrasekhar_hyro_1961, JULIEN_KNOBLOCH_1998, Stellmach_etal_2014_rbc_regimes}. At higher values of supercriticality, and for Prandtl numbers above unity, the flows transition to the convective-Taylor-column regime \citep{SPRAGUE_JULIEN_KNOBLOCH_WERNE_2006}. In this regime, the flow morphology is characterized by upflow and downflow columns each sheathed by counter-rotating fluid of the opposite sign of thermal fluctuation. Since these sheathed columns have zero net vorticity, they are non-interacting and freely drift throughout the domain. Thus, they do not organize into a stable geometric packing \citep{Grooms_2010_CTC, JulienGT2012, Stellmach_etal_2014_rbc_regimes}. At yet higher values of $\mathcal{R}$, the cellular structure is lost and the flows enter a plume regime characterized by the breakdown of vertically coherent columns \citep{JulienGT2012, Stellmach_etal_2014_rbc_regimes}. This breakdown occurs as the thermal boundary layer destabilizes and launches thermal plumes. Finally, at even higher values of supercriticality, vertical coherence breaks down further and the flows enter a geostrophic turbulence regime where pressure locally balances the Coriolis force \citep{Stellmach_etal_2014_rbc_regimes,Julien_GTHeat_2012PhysRevLett.109.254503, JulienGT2012}. 

There are several proposed models that aim to study moist convection at a fundamental level \citep[e.g.,][]{Pauluis2010_moistRBC_EOS, Hernandez-Duenas_Majda_Smith_Stechmann_2013, Smith2017_PQG, Chen2024_SPQG, VallisOishi2025}. The ``Rainy-Bénard" model from \citet{VPT2019} (hereafter VPT19) is a particularly simple example, designed to focus study on the buoyancy released from condensation. This model assumes that all condensed fluid coalesces into rain, falls out rapidly, and does not form clouds or experience re-evaporation. The set of equations from VPT19 has been used in several studies of moist convection \citep{VPT2019, OishiBrown24, Agasthya2024_cooledRainyBenard} which explored 2D domains without rotation. Prior studies of rotating moist convection used a quasi-geostrophic system \citet{Smith2017_PQG, Chen2024_SPQG} which assumes the limit of rapid rotation (small Rossby number). In this paper we present a suite of simulations of rotating moist convection without using a quasi-geostrophic system allowing us to study systems of more moderate rotation like Jupiter's weather layer. We use the the Rainy-Bénard model of VPT19 and discuss the effects of rotation on flow morphology and scaling laws of the heat transport and degree of turbulence. 

We explain the equations and our numerical methods in section \ref{sec:Methods}. We ran a suite of simulations across Rayleigh-Ekman space, and identify several morphological regimes which we discuss in section \ref{sec:morphology}. We then measure energy fluxes through the system, and report Nusselt and Reynolds number scalings in section \ref{sec:transport}. Finally, we discuss the implications of these results for further study of rapidly rotating planets such as Jupiter in section \ref{sec:discussion}.

\section{Methods} \label{sec:Methods}

        \begin{figure}
            \centering
            \includegraphics[width=0.95\columnwidth]{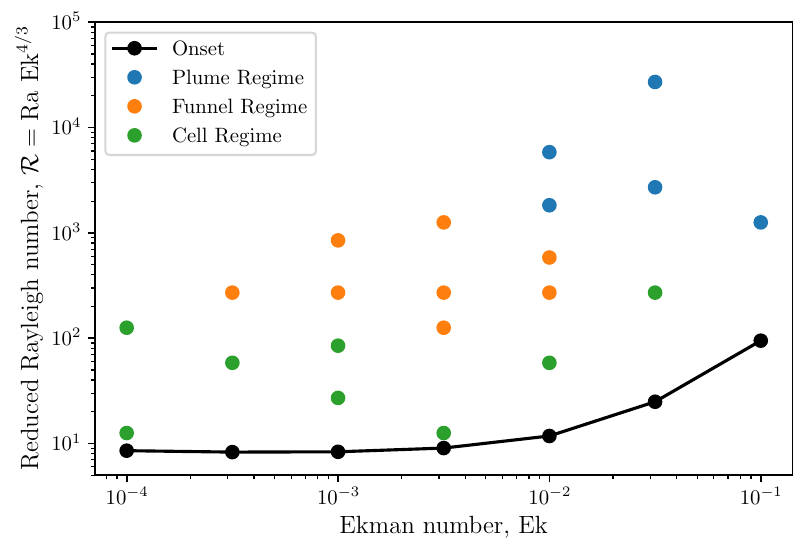}
            \caption{
            %The parameter space explored here in terms of Ekman number and reduced Rayleigh number ($\mathcal{R} \equiv \rm{Ra}\rm{Ek}^{4/3}$). Colored circles indicate parameters for which simulations were performed, with the color indicating the dynamical regime of each simulation. The black curve shows the marginal stability curve, i.e., the critical $\mathcal{R}$ for convective onset.
            The suite of simulations is shown on a plot of reduced Rayleigh number ($\mathcal{R} \equiv \rm{Ra}\,\rm{Ek}^{4/3}$) against Ekman number $\rm{Ek}$ as well as the critical reduced Rayleigh number curve in black. Note that if the Ekman number is changed while keeping $\mathcal{R}$ fixed, the free-fall time changes. The color of the points indicate the dynamical regime for each simulation.  
            }
            \label{fig:roadmap}
        \end{figure}
%\whitney{I'm just showing dimensional equations since I like expressing Nu in terms of dimensional quantities later. However, I don't feel too strongly about this and I could easily be convinced to change these to the non-dimensional equations}

 We seek to study how the localized heating from condensation interacts with changes in flow morphology arising from the Coriolis force. The Rainy-Bénard equations of VPT19 are well suited to this task since they simplify the system by neglecting the complex physics of the liquid phase, while still capturing the coupling between latent heat and buoyancy which is fundamental to this instability.  We add a Coriolis term appropriate for a Cartesian geometry (with spatial variables $x$, $y$, and $z$, time $t$, and coordinate unit vectors $\hat{\mathbf{x}}$, $\hat{\mathbf{y}}$, and $\hat{\mathbf{z}}$) under the polar $f$-plane approximation---i.e., the rotation vector is spatially uniform and aligned with gravity. These equations, in dimensional form, are
\begin{eqnarray}
    \frac{D\mathbf{u}}{Dt} &=& -\nabla\phi + g\alpha\theta \hat{z} + \nu {\nabla}^2 \mathbf{u} - 2 \mathbf{\Omega}\times\mathbf{u} \;\label{eqn:u} ,
\\
    \frac{D\theta}{Dt} &=&\frac{\Lambda-\Lambda_s}{\tau}\mathcal{H}(\Lambda-\Lambda_s)+\kappa{\nabla}^2\theta \; \label{eqn:theta},
\\
    \frac{D\Lambda}{Dt} &=& - \frac{\Lambda-\Lambda_s}{\tau}\mathcal{H}(\Lambda-\Lambda_s) + \kappa_m {\nabla}^2 \Lambda \;\label{eqn:q} ,
\\
    \nabla\cdot\mathbf{u} &=& 0 \; ,
\end{eqnarray}

\noindent where $\frac{Df}{Dt} = \frac{\partial f}{\partial t} + \mathbf{u}\cdot\nabla f $ is the material derivative and $\mathbf{u} = u \hat{\mathbf{x}}+v \hat{\mathbf{y}} + w \hat{\mathbf{z}}$ is the velocity vector. We define the latent heat associated with the specific humidity of a parcel as  $\Lambda \equiv \frac{L}{c_P}q$ where $L$ is the latent heat of condensation, $c_P$ is the specific heat at constant pressure, and the specific humidity, $q$, measures the mass of condensable vapor per unit mass of fluid. Here $\Lambda$ has units of temperature and represents the change in temperature should all of the vapor phase condense out, and $\Lambda_s$ is the latent heat evaluated for a fully saturated parcel . The potential temperature, $\theta$, is linearly related to the fluid temperature $T$, via
\begin{equation}
    T=\theta - \frac{g}{c_p}z .
\end{equation}
The constant $g$ is the gravitational acceleration ($\mathbf{g} = -g \hat{\mathbf{z}}$) and $\alpha$ is the coefficient of thermal expansion. We can further define the moist static energy 
\begin{equation}
    m = \theta + \Lambda
\end{equation}
which here has units of temperature and represents the thermal and latent heat energy associated with a fluid parcel. The gradient of $m$ determines the stability of this system. The rotation vector, $\mathbf{\Omega} = \Omega \hat{z}$, is anti-parallel to gravity. The diffusivities $\nu$, $\kappa$, and $\kappa_m$ are the viscosity, thermal diffusivity, and moisture diffusivity, respectively. The constant $\tau$ is the condensation time scale and $\mathcal{H}$ is a smooth step function given by $\mathcal{H}(\Lambda-\Lambda_s)=\frac{1}{2}[1 + \mathrm{tanh}(k(\Lambda-\Lambda_s))]$ where we take $k=1.9 \times 10^4$  following VPT19. This functional form ensures that moisture condenses rapidly whenever the specific humidity exceeds the saturation level $\Lambda_s$.

Following VPT19, we take the specific humidity at saturation to be  
\begin{equation}
    \Lambda_s(T)=\Lambda_0 e^{A T},
\end{equation}
with constants $\Lambda_0$ and $A$, as an approximation of the solution to the Clausius–Clapeyron equation for an ideal gas with the assumption that the moisture vapor pressure is much less than atmospheric pressure. This stratification in humidity leads to an up-down asymmetry in the system, unlike the traditional (dry) case of Rayleigh-B\'enard convection. An adiabatically rising parcel of fluid will cool with height, causing the saturation value of specific humidity to fall. If this parcel is fully saturated, it will become supersaturated as it rises, causing condensation of the excess moisture. This releases latent heat, raising the buoyancy and increasing the parcel's upward acceleration. Thus, condensation in upflows creates a feedback loop that enhances the speed of buoyant rise. An equivalent instability does not exist for downflows, since as a fluid parcel falls the parcel becomes increasingly subsaturated. No evaporation occurs in the bulk since the Rainy-B\'enard equations assume that all liquid phase water rapidly leaves the box as rain and thus the feedback loop is not triggered.

Our simulations have a vertical extent of $H$ and span three moisture scale heights (i.e., $AgH/c_P = 3$ which corresponds to $\alpha_{\mathrm{VPT}} =3$ in the notation of VPT19). We use a periodic domain in $x$ and $y$ with aspect ratio 10 such that the horizontal extent of our box is $10H$ in both $x$ and $y$. We adopt impenetrable, no-slip, fixed-temperature, and fully saturated boundary conditions, 

\begin{eqnarray}
    &&\mathbf{u}(z=0) = \mathbf{u}(z=H) = 0 \; ,
\\
    &&\theta(z=0)=\theta(z=H) = 0 \; ,
\\
    &&\Lambda(z=0) =\Lambda_s(z=0)= 0.19 \;, 
\\    
    &&\Lambda(z=H) =\Lambda_s(z=H)=  0.19e^{-3} \; .
\end{eqnarray}

We initialize our simulations with a fully saturated moisture profile ($\Lambda = \Lambda_s$) and zero velocity ($\mathbf{u} = 0$). The initial condition for potential temperature is an isothermal profile ($\theta=0$) plus a Gaussian noise field. The noise field has an amplitude of $M = 10^{-1}$ and is multiplied by a sinusoid with a half-wavelength equal to $H$ such that the noise field is zero at the upper and lower boundaries to ensure that the initial conditions do not violate the boundary conditions ($\theta = M~ \mathrm{sin}(z/\pi) G$ where G is a Gaussian noise field). 

%This differs from VPT19 which initialized the simulations with a localized supersaturated perturbation.

%\subsection{Nondimensional Numbers}
%\brad{Perhaps you discuss the boundary conditions here.  Then start a new section that discussed the nondimensional numbers. Start with the control parameters Ra, Pr, etc. then discuss the output numbers (Re, Nu, etc.)  You might consider defining the compensated Rayleigh number here and give it a symbol (perhaps \cal R.  You refer to the compensated Raleigh number so often that a symbol is probably useful.}

Our simulations are all performed using a nondimensional form of these fluid equations, the details of which are discussed in Appendix \ref{sec:ApdxA}. The behavior of this system of equations is characterized by six nondimensional control parameters: the moist Rayleigh number ${\rm Ra}_m$, Prandtl number ${\rm Pr}$, Lewis number ${\rm Le}$, Ekman number ${\rm Ek}$, nondimensional temperature scale for saturation $\widetilde{A}$, nondimensional adiabatic lapse rate $\widetilde{B}$, and a nondimensional condensation timescale $\widetilde{\tau}$, each given by

\begin{eqnarray}
    \mathrm{Ra}_m \equiv \frac{t_{ff}}{t_{therm}t_{visc}} &=& \frac{g\alpha\Delta m H^3}{\kappa\nu} = \frac{g\alpha(\Delta \theta + \Delta \Lambda)H^3}{\kappa\nu} \; ,
\\
    {\rm Pr} \equiv \frac{\nu}{\kappa} \;, \qquad {\rm Le} &\equiv& \frac{\kappa_m}{\kappa}\; , \qquad {\rm Ek} \equiv \frac{\nu}{2\Omega H^2} \; ,
\\
    \widetilde{A} \equiv A\frac{g\alpha}{\Delta m}\;, \qquad\widetilde{B} &\equiv& \frac{g^2\alpha H}{c_p\Delta m}\;,
 \qquad \widetilde{\tau} \equiv \frac{\tau}{t_D} \; ,
\end{eqnarray}  
where $t_D=H^2/\kappa$ is a thermal diffusion time (see Appendix \ref{sec:ApdxA}). To capture buoyancy generation by condensation in our definition of the moist Rayleigh number, we use the moist static energy from VPT19, $m \equiv \theta+\Lambda$, and an associated free-fall time $t_{ff} = \sqrt{H/(g\alpha\Delta m)}$.  For our freefall time we evaluate $\Delta \theta$, $\Delta \Lambda$, and $\Delta m$ across the domain with values dictated by our boundary conditions. Constants $\widetilde{A}$ and $\widetilde{B}$ are similar to the constants $\alpha$ and $\beta$ from VPT19, but reflect the differences in equation formulation and nondimensionalization.

Note that this moist Rayleigh number is the sum of the standard, `thermal' Rayleigh number from Rayleigh-B\'enard convection and a contribution from moisture. Our moist Rayleigh number thus reduces to the standard Rayleigh number of Rayleigh-B\'enard convection when $\Delta \Lambda=0$. For our choice of an initial isothermal stratification with fixed-temperature boundary conditions, the potential temperature jump across the layer vanishes ($\Delta\theta = 0$), and thus the thermal Rayleigh number is zero, i.e., the system would be neutrally stable without condensation. Note also that, instead of the Rayleigh number, we will often find it convenient to employ the reduced Rayleigh number, $\mathcal{R} = \mathrm{Ra}_m \, \mathrm{Ek}^{-4/3}$, which represents the scaling of the critical Rayleigh number with rotation in the rapidly rotating regime.

We set all of the diffusivities to the same value; therefore, the Prandtl and Lewis numbers are unity, $\rm{Pr}={\rm Le}=1$. Further, we take an extremely fast condensation timescale relative to the diffusive timescales: $\widetilde{\tau} = 5\times10^{-5} $.

%We set our potential temperature to be the same at the bottom and top of the domain $\theta(z=0)=\theta(z=1) = 0$. Our domain spans three moisture scale heights, and our moisture boundary conditions are $q(z=0)=1$ and $q(z=H) = e^{-3}$. 

%\subsection{Numerical methods} 
%We solve for convective onset with a numerical eigenvalue solver using the pseudospectral tau method implemented in Dedalus v3 \citep{Burns_dedalus_2020PhRvR...2b3068B}. The methods for linearizing the Rainy-Bénard equations and implementing an eigenvalue solver are discussed in Ref.~\citep{OishiBrown24}. 

We solve for convective onset with a numerical eigenvalue solver using the pseudospectral tau method implemented in Dedalus v3 \citep{Burns_dedalus_2020PhRvR...2b3068B}. The methods for linearizing the Rainy-Bénard equations and implementing an eigenvalue solver are discussed in \citet{OishiBrown24}. Our direct numerical simulations are implemented in Dedalus v2 using the third-order semi-implicit BDF timestepping scheme (``SBDF3") given by Eq.~(2.14) of \citet{Wang_timesteppers_2008} with time step size set by an advective Courant-Friedrichs-Lewy condition with safety factor $0.3$. The time-step size is also not allowed to exceed $0.1\tau$ to ensure that condensation is adequately resolved. Simulations are run until volume averaged outputs such as Reynolds number and Nusselt number converge to a statistically steady state. 

\section{Morphological Regimes} \label{sec:morphology}
We measure the critical Rayleigh number for each value of Ekman number used in this study. The onset curve is shown by the black line in Fig.~\ref{fig:roadmap}. Consistent with the dry case \citep{Chandrasekhar_hyro_1961}, we find that convective onset occurs at a fixed value of the reduced Raleigh number (${\cal R} \approx 8$) in the rapidly rotating regime $\rm{Ek} \lesssim  10^{-3}$. With our chosen parameters, the system would be stable in the absence of moisture (i.e., the thermal Rayleigh number is zero). With an unstable thermal gradient, the onset of dry Rayleigh B\'enard convection in the rapidly rotating regime occurs at a similar value of $\mathcal{R}\approx 7$ \citep{Chandrasekhar_hyro_1961}.

We ran a suite of simulations where we vary the Ekman and Rayleigh numbers. Each simulation is indicated by a point in Fig.~\ref{fig:roadmap} and we include a table of simulations in appendix \ref{sec:ApdxB}. We observe three dynamical regimes in our simulations (cells, funnels, and plumes) and indicate the location in parameter space where each dynamical regime exists in Figure \ref{fig:roadmap} with the color of the points. The cellular regime, indicated by the green points in Fig. \ref{fig:roadmap}, is found at reduced Rayleigh numbers $\mathcal{R}$ just above onset. At approximately $\mathcal{R} \gtrsim 100$, and with a sufficiently low Ekman number, the flow transitions to the funnel regime. At yet higher values of reduced Rayleigh number, $\mathcal{R}\gtrsim 10^3$, we observe the plume regime.

\subsection{Cells}
\begin{figure}
    
            \centering
            \includegraphics[width=0.95\columnwidth]{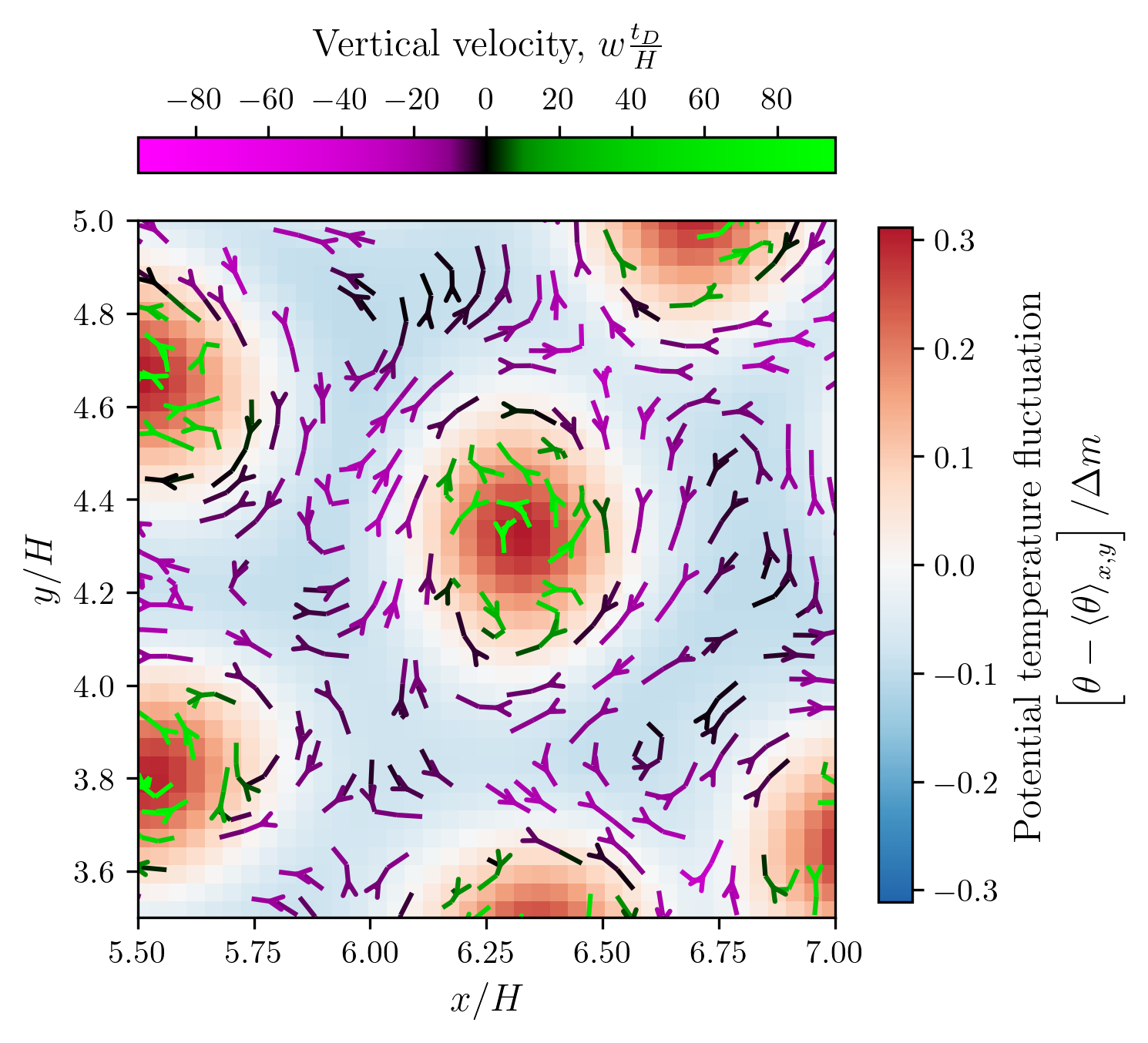}
            \caption{Velocity stream lines for a single cell are shown over a plot of potential temperature fluctuations about the horizontal mean at $z=0.8\, H$. The color of the stream lines indicates the vertical velocity $w$. The streamlines have a cyclonic vorticity at the center of the cell and at the periphery of cell the flows are anti-cyclonic. There are additionally cyclonic vortexes in the downflowing cell vertices.}
            \label{fig:cells_stream}
        \end{figure}

\begin{figure}
            \centering
            \includegraphics[width=0.95\columnwidth]{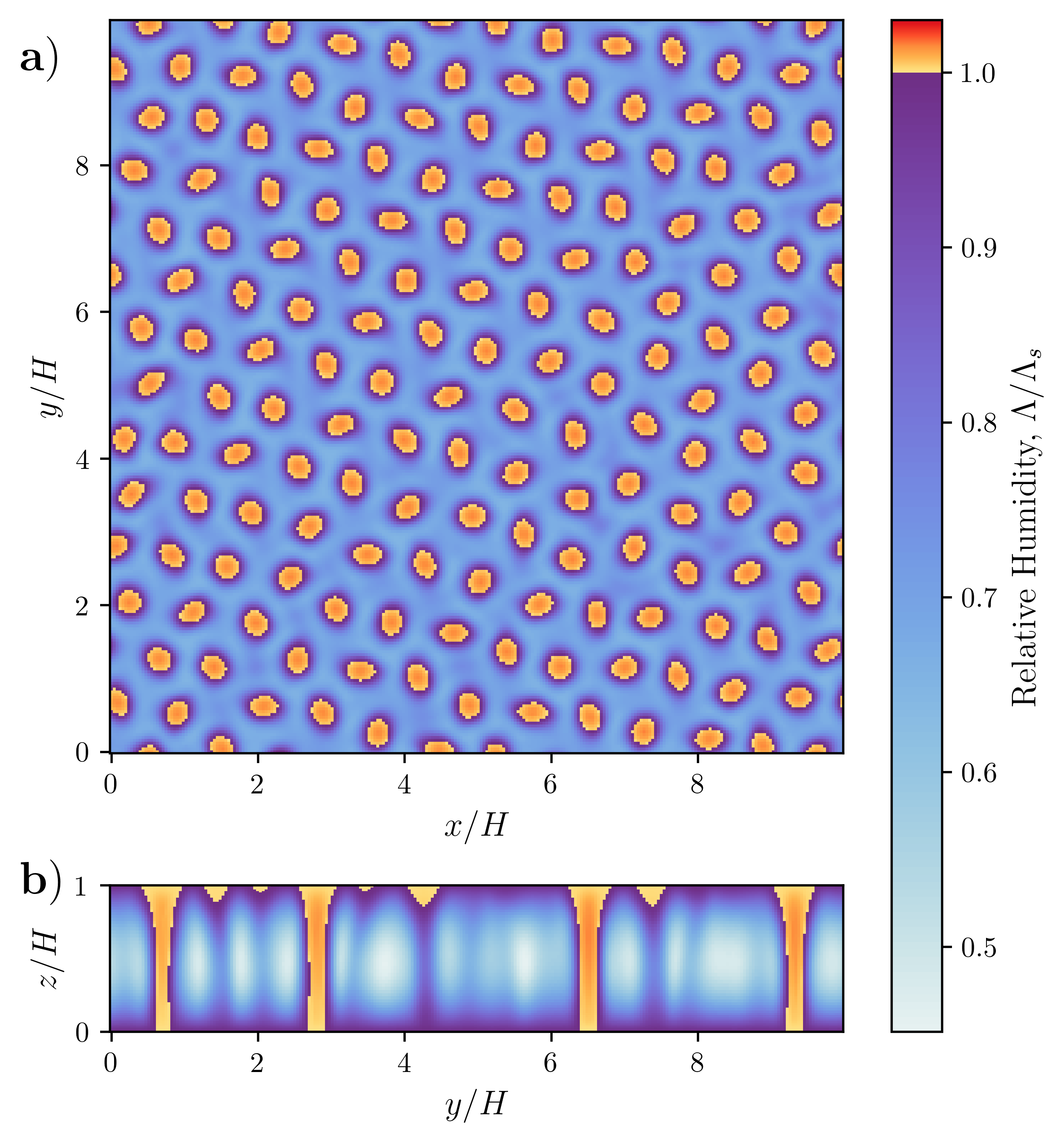}
            \caption{The cell-regime dynamics are shown in plots of relative humidity ($\Lambda/\Lambda_s$). Panel a shows a horizontal slice at $z=0.8\, H$, and hexagonally packed, supersaturated upflow cells can be seen. Supersaturated regions are marked by the discontinuity in the colorbar. In panel b we show a vertical slice at $x=0$, where we see that the supersaturated columns do not contain any inclusions and extend over the entire vertical domain. The center of the cell is cyclonic, which is surrounded by anticyclonic flows.}
            \label{fig:cells}
        \end{figure}

\begin{figure*}
    \centering
    \includegraphics[width=0.95\textwidth]{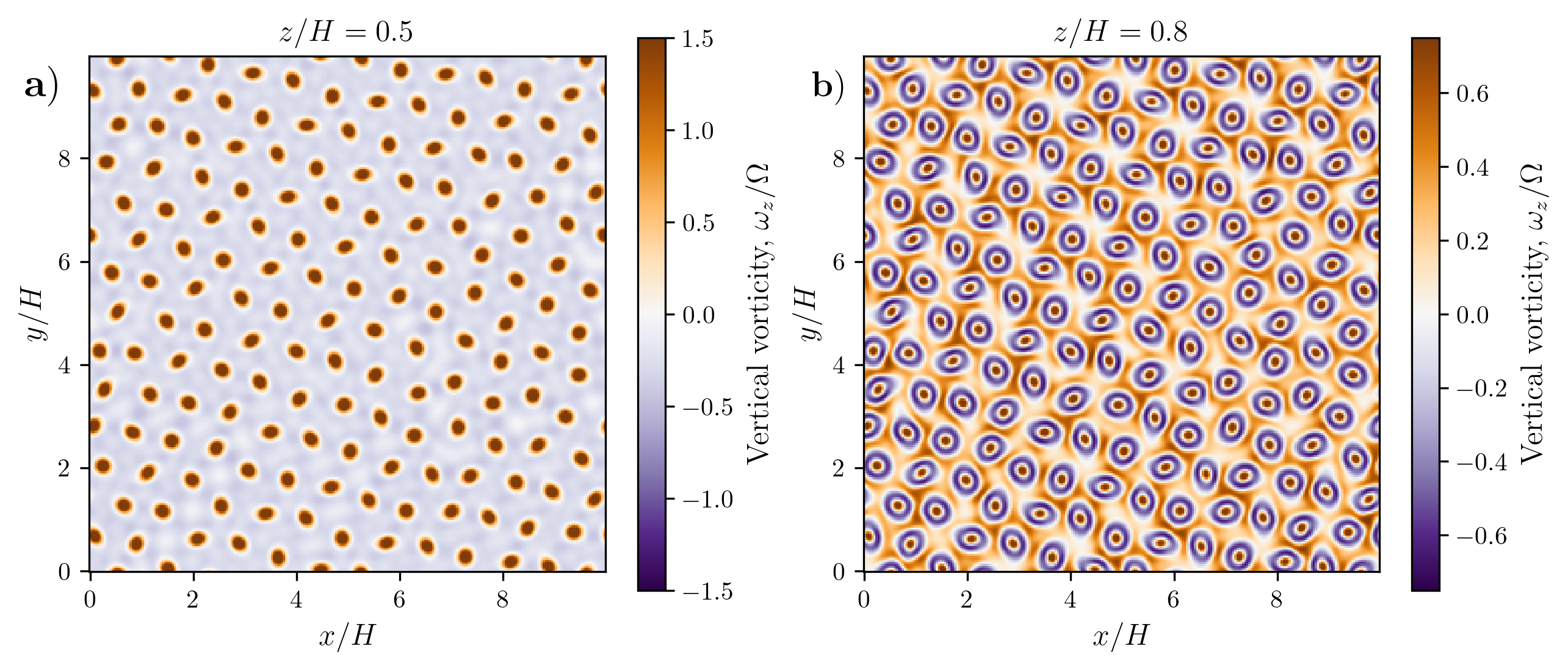}
    \caption{Vertical vorticity $\omega_z$ scaled by the rotation rate $\Omega$ at heights $z=0.5 \, H$ and $z=0.8 \, H$ for flows in the cell regime. At the midplane (panel a), upflows form compact cyclonic columns and are surrounded by diffuse anticyclonic downflows. At the top of the domain (panel b) an annular structure develops. The center of the upflow columns has positive (cyclonic) vorticity, which is surrounded by a ring of anticyclonic fluid, which is then surrounded by cyclonic downflows. }
    \label{fig:cell_vorticity}
\end{figure*}

For weakly supercritical values of $\mathcal{R}$, we observe cellular dynamics. This cell regime, %exemplified by the simulation with $\rm{Ra}_m = 2.7 \times 10^5$, $\mathcal{R} = 2.7\times 10^1$ and $\rm{Ek}=10^{-3}$ in Figures \ref{fig:cells_stream}-\ref{fig:cells}, 
demonstrated in Figs.~\ref{fig:cells_stream}-\ref{fig:cell_vorticity} for the case with $\rm{Ra}_m = 2.7 \times 10^5$ and $\rm{Ek}=10^{-3}$ ($\mathcal{R} = 27$),
is characterized by spinning columns of hot, supersaturated upflows embedded within a network of cold, dry downflows (see Fig. \ref{fig:cells_stream}). The upflows organize into a regularly packed hexagonal structure (see Fig. \ref{fig:cells}). As in dry rotating convection \citep{Stellmach_etal_2014_rbc_regimes}, the upflows are typically cyclonic in the bottom of the domain where the horizontal component of the flow is convergent and anticyclonic at the top where the horizontal flows are divergent.  Conversely, the downflow network is cyclonic at the top and anticyclonic at the bottom. Unlike dry Rayleigh-B\'enard convection, the upflows and downflows are not symmetric; the upflows form compact columns that are buoyantly driven by latent heat release. Each upflow column is fully saturated across the whole domain, thus latent heat is released continuously as a fluid parcel rises, accelerating its ascent. In contrast, the downflows are relatively passive with no equivalent buoyancy source or sink to accelerate a down-welling parcel's descent.

As the reduced Rayleigh number increases above marginal values, an additional feature unobserved in rotating Rayleigh-B\'enard convection develops. The vertical vorticity in the upper portion of the domain develops a ringed structure in the upflow regions (see Fig. \ref{fig:cell_vorticity}), with an inner core of cyclonic flow sheathed by a ring of anticyclonic flow which in turn is surrounded by the cyclonic downflow network. Vorticity is larger near the bottom of the domain and the mid z-plane than at the top of the domain. The annular vorticity structures may be a signature of a transition to the funnel regime.

%\brad{In my mind, one of the primary features of this cellular regime is the vertical coherence.  You might consider adding a paragraph about how the downflows and upflows extend from top to bottom as a coherent structure.  Further, you should layout how these flows differ from the {\bf cellular regime} in rotating Rayleigh-Benard.  To wit, the upflows and downflows are NOT symmetric (as they are in Rayleigh-Benard convection). The latent heat release accelerates the upflows but is essentially passive in the downflows.  Hence, there is an up-down asymmetry that is reminiscent of convection in stratified fluids.  The upflows are compact, while the downflows form an interconnected network. The net structure of upflows and downflows has a roughly hexagonal pattern.  You might also mention how the vorticity behaves. I expect that the upflows are cyclonic in the upper portion of the domain and anticyclonic in the lower portion. This is consistent with an upflow having to diverge as it approaches the upper surface (anti-cyclonic) and with an upflow drawing in converging fluid at the bottom (cyclonic).  So the spin of an upflow column must change direction somewhere in the middle of the domain.}

\subsection{Funnels}
\begin{figure}
            \centering
            \includegraphics[width=0.95\columnwidth]{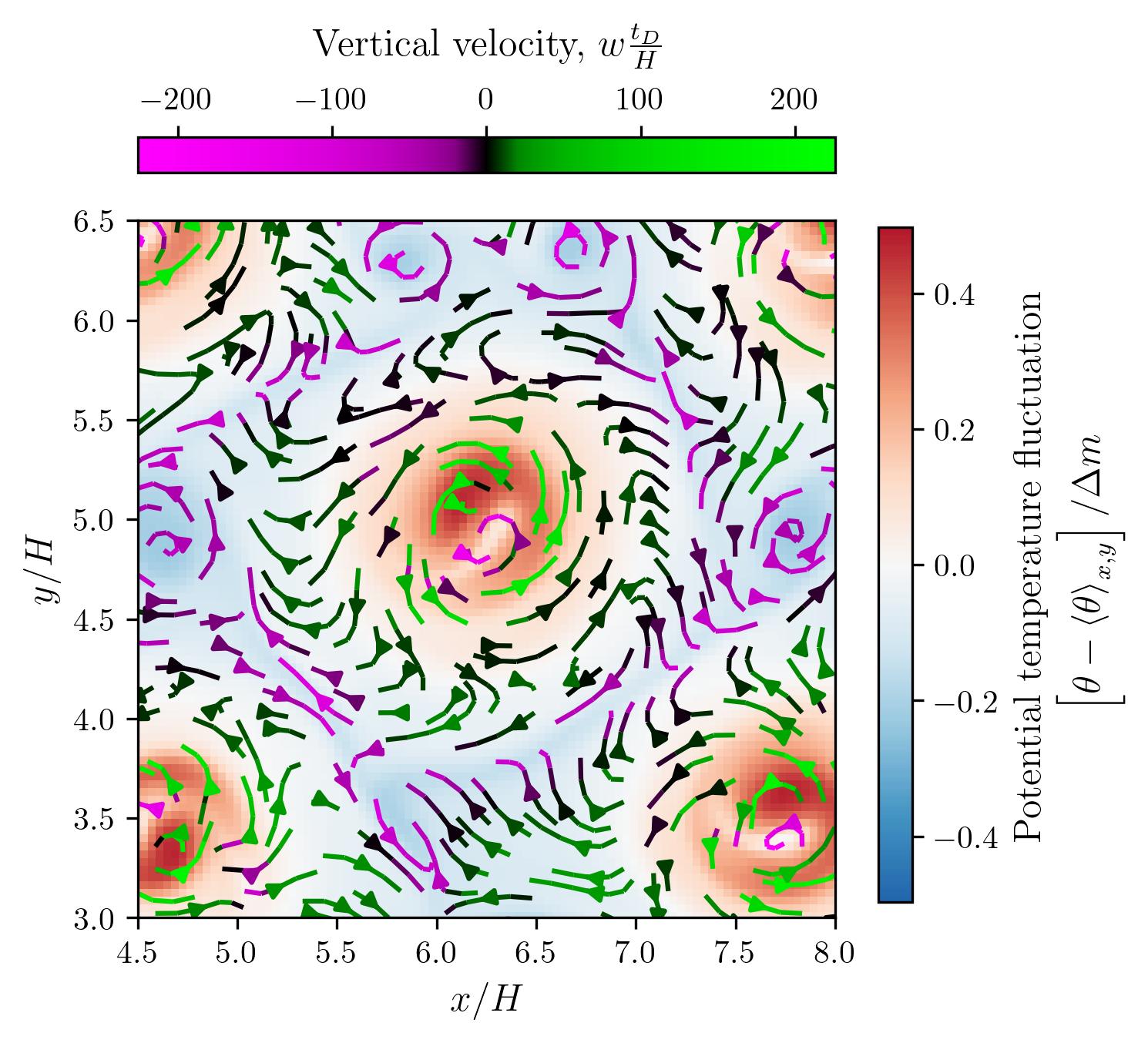}
            \caption{Velocity stream lines for a single funnel is shown over a plot of potential temperature fluctuations about the horizontal mean at $z=0.8\,H$. The color of the stream lines indicates the vertical velocity $w$. We observe counter-clockwise rotating upflows diverging from the upflow column. Streamlines cross the spiral core and descend while crossing this feature. The cell boundaries form clockwise vortexes of descending fluid at the vertices of the hexagonal cells.}
            \label{fig:funnel_cells_stream}
        \end{figure}

\begin{figure}
            \centering
            \includegraphics[width=0.95\columnwidth]{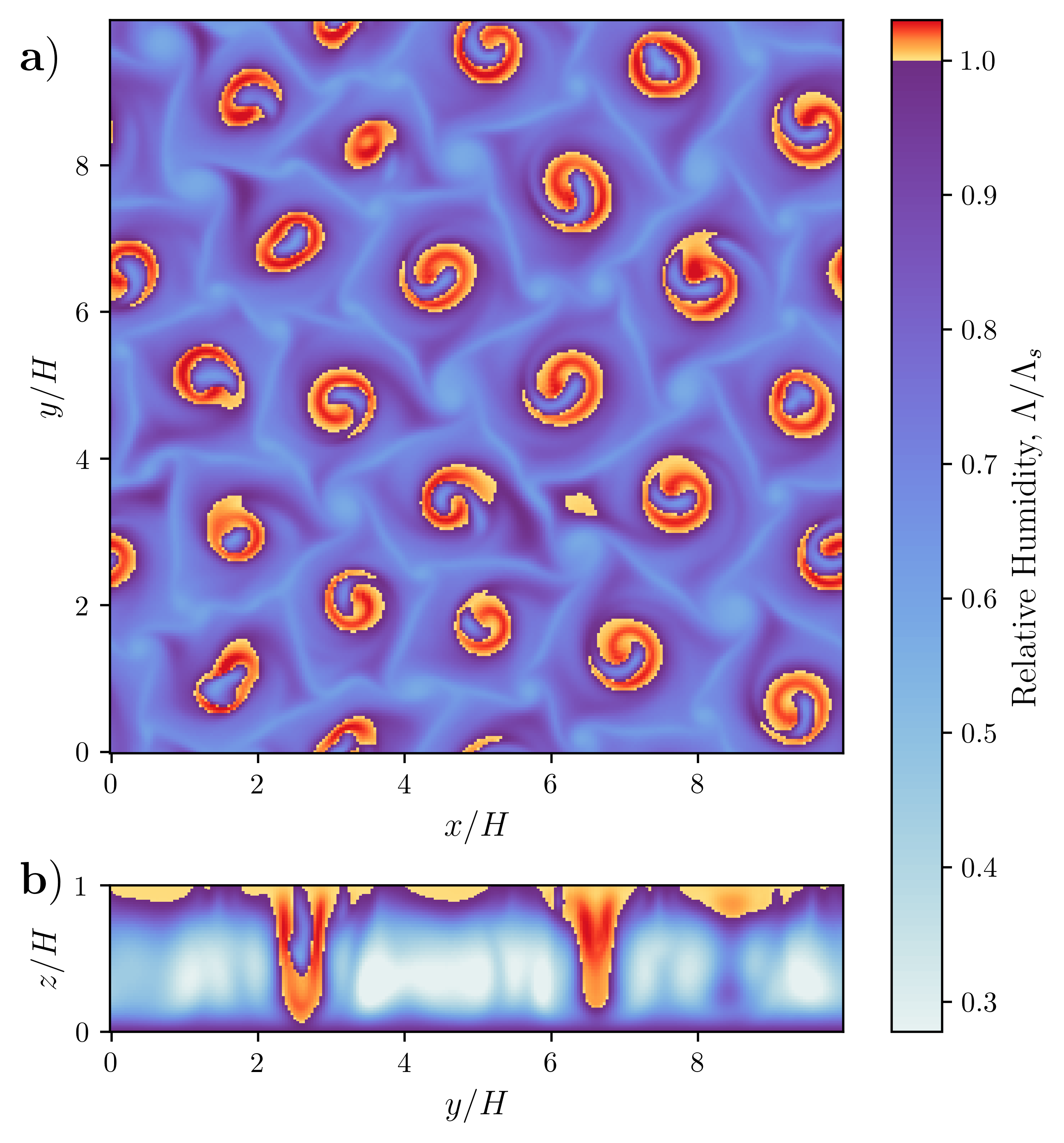}
            \caption{ The funnel regime dynamics are shown in plots of relative humidity. Panel a shows a horizontal slice at $z = 0.8\,H$. Supersatuated regions (indicated by shades of blue) form spirals or rings with an inner core of dry fluid. These funnels have a regular hexagonal packing, and vortex structures can be seen at the vertices of the hexagons. In panel b we show a vertical slice at $x = 0$, where we see the funnel-shaped structure. The supersaturated regions extend across the domain, expanding horizontally as they rise. At approximately $z=0.5\, H$ an inner core of dry fluid emerges, and this inner core extends to the top of the domain. }
            \label{fig:funnel_cells}
        \end{figure}

\begin{figure*}
            \centering
            \includegraphics[width=0.95\textwidth]{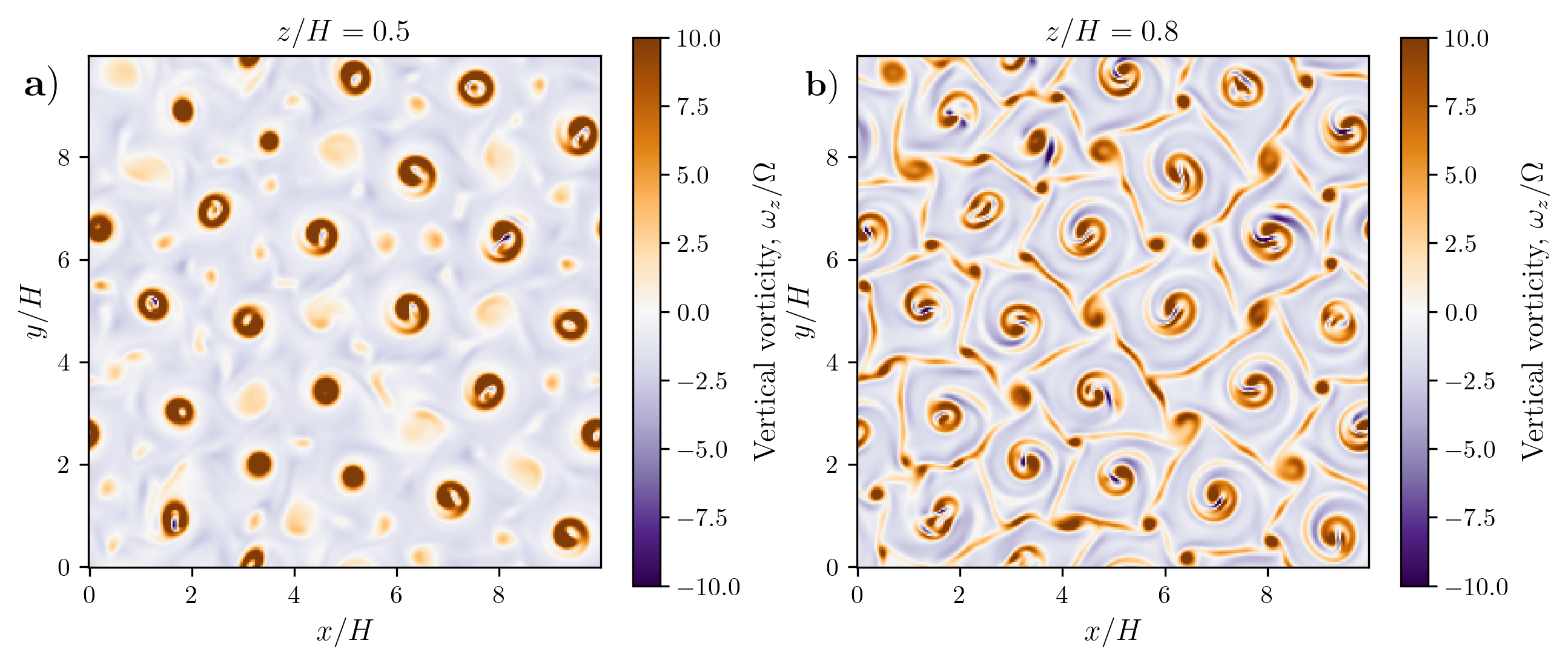}
            \caption{Vorticity structures for a funnel-regime simulation at $z=0.5\,H$ and $z=0.\,H8$. At the mid-$z$ plane (panel a), upflow columns have cyclonic vorticity with an inner core with lower vorticity emerging in some cells, either as a fully enclosed region or as a spiral arm connecting the inner core to the anticyclonic downflows surrounding the upflow columns. At $z=0.8\,H$ (panel b) the upflow columns have destabilized, leading to spiral arms of cyclonic and anticyclonic flow to develop at the center of each cell. The spiral structure is surrounded by anticyclonic flow, and the boundaries between each cell are cyclonic. }
            \label{fig:funnel_vorticity}
        \end{figure*}

The funnel regime, like the cell regime, is characterized by cells with a hexagonal packing, with supersaturated columns of hot, rising flow, and cell boundaries with cold, dry, descending flows. However, embedded within each ascending column is an inner spiral core of cold, dry, descending fluid. This inclusion extends from the top of the domain to approximately the middle of the domain. This flow structure can be seen in the simulation with $\rm{Ra}_m=5.8 \times 10^5$ and $\rm{Ek}=3.16\times10^{-3}$ ($\mathcal{R} = 270$) shown in Figures \ref{fig:funnel_cells_stream}-\ref{fig:funnel_vorticity}. In the lower portion of the domain, upflows form columns of cyclonic vorticity. The structure begins to change in the middle of the domain ($z=0.5$) with an inner core of colder, drier, less cyclonic fluid developing within each upflow column. In the upper portion of the domain this new structure continues to develop and forms spirals of intertwined cyclonic and anticyclonic flows in each column. The streamlines from the hot upflow regions then diverge and flow into cyclonic downflows at the vertices of the hexagonal cells. At high values of reduced Rayleigh number, inner spiral fragments and the structure takes on a more turbulent appearance.
%\brad{When you first started explore these funnel cells, I too believed that the funnel cells could be a transitional regime.  However, as you have explored more of the parameter space, it looks to me like its a regime of its own.  You don't really see a continuum of behavior.  Do you feel differently?}


    \subsection{Plumes}
\begin{figure*}
            \centering
            \includegraphics[width=0.95\textwidth]{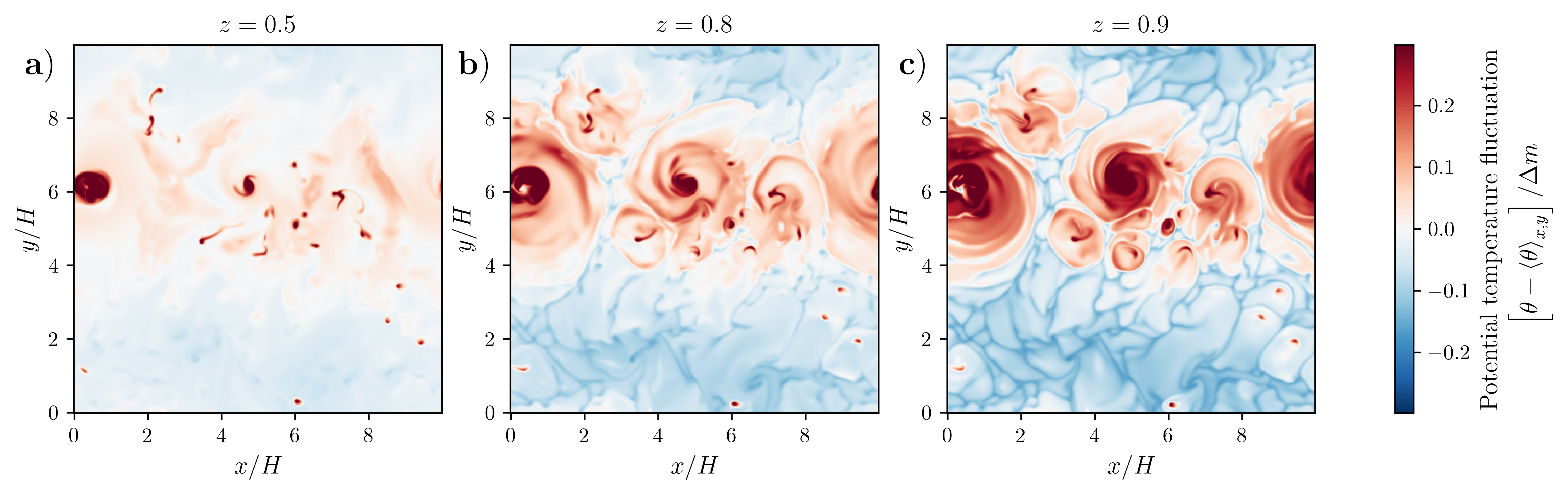}
            \caption{Horizontal slices of potential temperature fluctuations about the horizontal mean in the plume regime are shown at $z=0.5\,H$ (panel a), $z=0.8\,H$ (panel b), and $z=0.9\,H$ (panel c). At $z=0.5\,H$ flows form narrow upflow structures. These upflows happen to aggregate and cluster in a line at $y=6 \, H$. At $z=0.8\, H$ we see granular structures start to emerge, and are most apparent in the upflow at $y=6\,H$. At $z=0.9\, H$ the granular structure is most clear and can be seen in both the upflows at $y=6\, H$ and the downflow lane at $y=2\,H$. }
            \label{fig:plume_buoyancy}
        \end{figure*}

\begin{figure}
        \centering 
        \includegraphics[width=0.95\columnwidth]{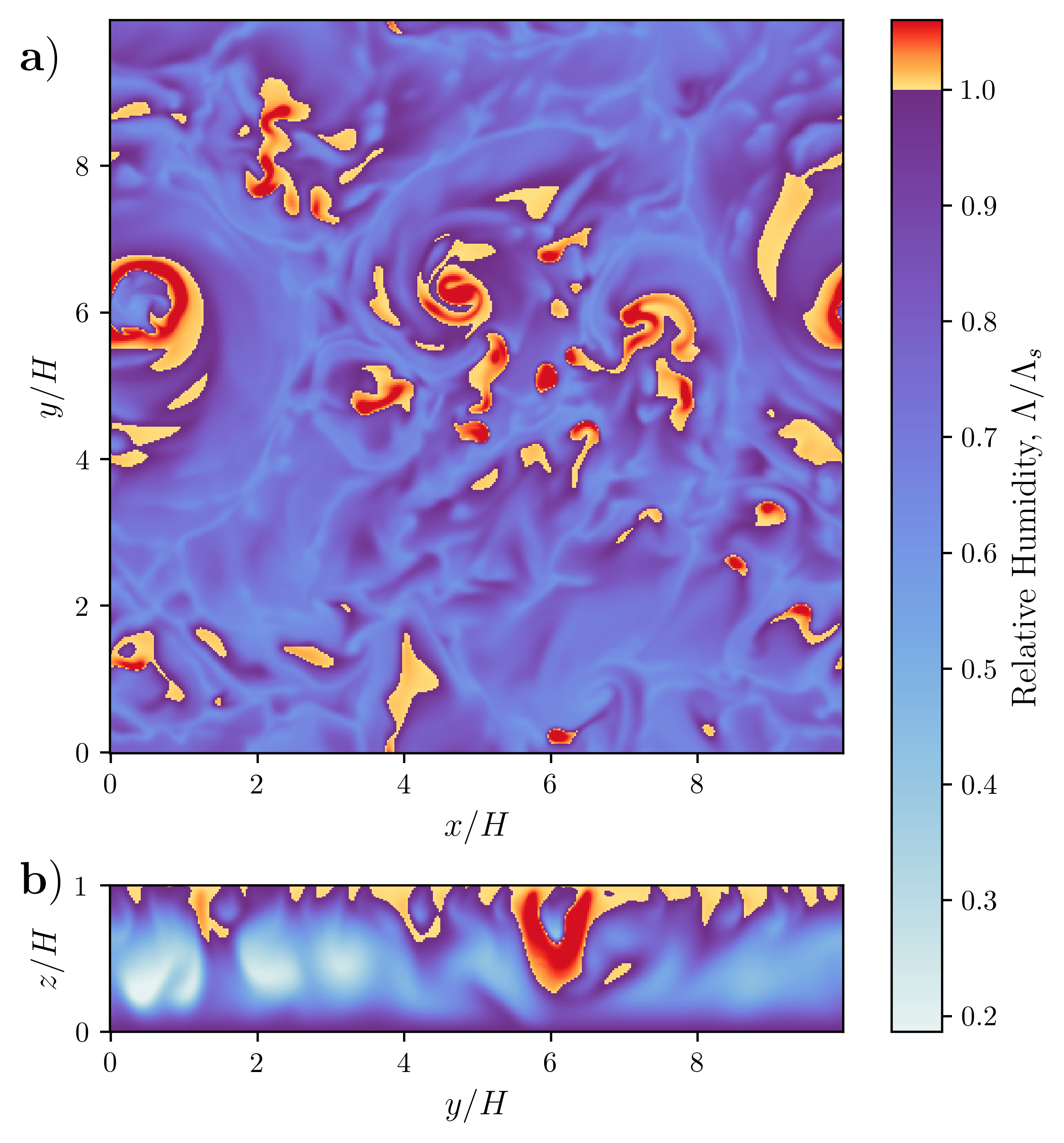}
            \caption{The plume-regime dynamics are shown in plots of relative humidity. Panel a shows a horizontal slice at $z = 0.8\,H$. No regular cell-like pattern is apparent. Supersaturated regions are concentrated in a line at approximately $y=6\,H$
            In panel b we show a vertical slice at $x = 0$. Here we see a strong plume with a funnel-like shape which extends across most of the domain at $y=6\,H$. We also see smaller plumes of supersaturated fluid form at approximately $z>0.8\,H$.}
            \label{fig:plume_RH}
        \end{figure}

\begin{figure*}
            \centering
            \includegraphics[width=0.95\textwidth]{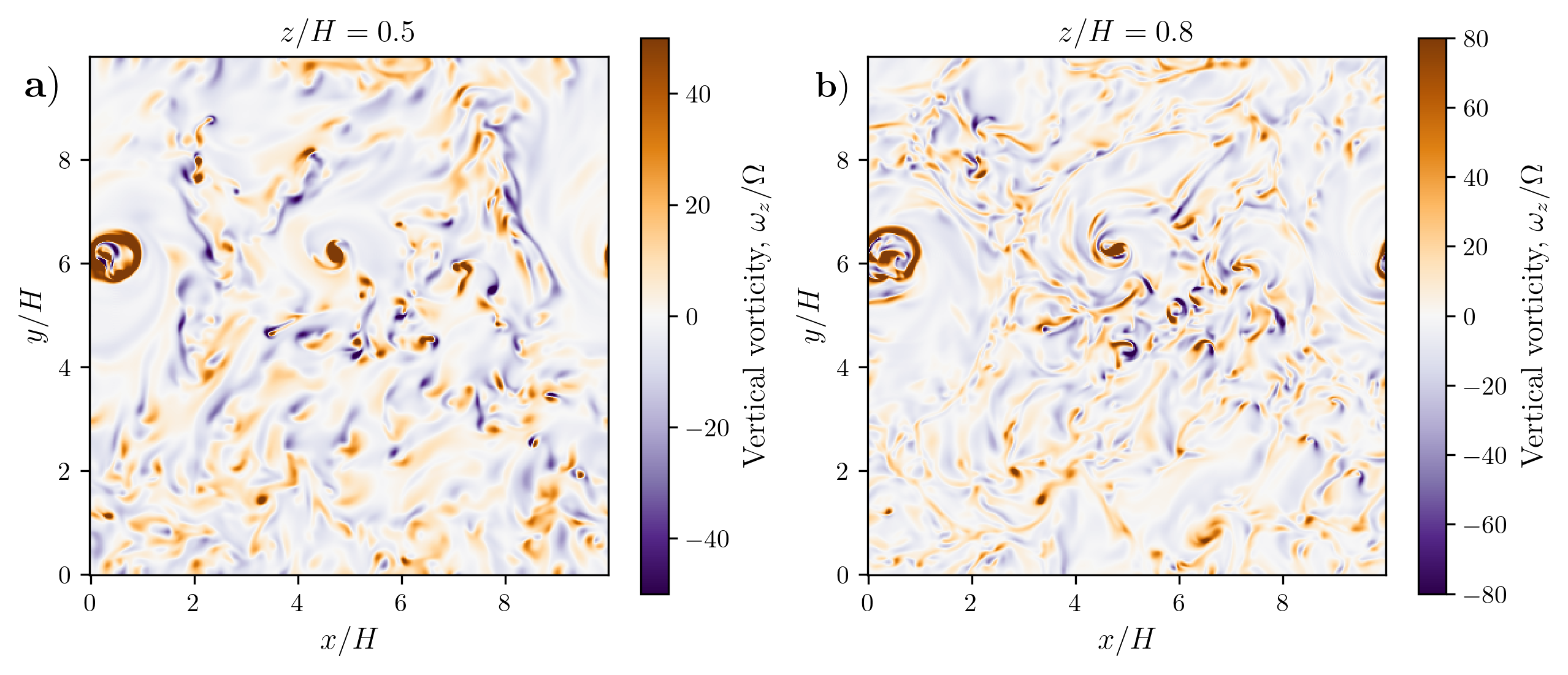}
            \caption{Horizontal slices of the $z$ component of vorticity in the plume regime are shown at $z=0.5\,H$ (panel a) and $z=0.8\,H$ (panel b). The vorticity structures appear similar at both depths with the most prominent features being strong cyclonic upflows.  }
            \label{fig:plume_vorticity}
        \end{figure*}

    The plume regime, with a characteristic simulation with $\mathrm{Ra}_m=2.71\times10^6$ and $\mathrm{Ek}=10^{-2}$ displayed in Figs \ref{fig:plume_buoyancy}-\ref{fig:plume_vorticity} is typified by less ordered flow patterns.  Upflow plumes at middle depths are narrow and merge over time. Near the top of the domain more plumes branch off, forming irregularly shaped granules. Signatures of the flows in the midplane persist in the upper boundary, with regions above the upflows forming hot-spots and regions above the downflows forming cold regions. The flows do not organize into a regular geometric packing. Signatures of rotation are apparent, especially in upflow plumes such as the one at $x=5$, $y=6$, however the overall structure is reminiscent of non-rotating flows. This regime is similar to the plume regime of rotating Rayleigh-B\'enard convection with the primary difference emerging from the up-down asymmetry inherent to the moist convection instability. The up-down asymmetry manifests clearly in the mid-$z$ plane, where upflows form compact structures with strong thermal fluctuations, and downflows are diffuse and have a weaker thermal fluctuation. The upflows are continuously driven by latent heating after they launch from the lower boundary, whereas once the downflows descend from the upper boundary they behave passively as compared to the upflows as there is no buoyancy sink analogous to the buoyancy source from condensation.
     

%\subsection{Scalings}
\section{Energy Transport} \label{sec:transport}
Nusselt number scalings are a frequently used method to quantify trends in convective heat transport. The Nusselt number of some scalar quantity, usually heat is the ratio of the total flux to the diffusive flux of that scalar quantity. In studies of dry, incompressible convection with stress-free boundary conditions the heat flux Nusselt number scales with reduced Rayleigh number as $\rm{Nu}\propto \mathcal{R} ^{3/2} $  \citep{Julien_GTHeat_2012PhysRevLett.109.254503, Stellmach_etal_2014_rbc_regimes}. In simulations with no-slip boundary conditions the Nusselt number scaling diverges from the asymptotic prediction and follows a $\rm{Nu} \propto \mathcal{R}^{3} $ scaling at lower supercriticality before converging towards a $\rm{Nu}\propto \mathcal{R} ^{3/2} $ scaling \citep{Stellmach_etal_2014_rbc_regimes} as the reduced Rayleigh number becomes sufficiently large. When considering a set of simulations at fixed Rayleigh number that are rotating slowly (${\rm Ro} \gg 1$), the Nusselt number is approximately constant as the Rossby number is varied. In the limit of rapid rotation (${\rm Ro} \ll 1$) at fixed Ra, the Nusselt number trends towards unity as convective motions are suppressed by rotation. At sufficiently large values of Raleigh number there is a clear transition between these two regimes and for Prandtl numbers near unity there is a slight enhancement in Nusselt number within this transition. At higher values of Prandtl number this enhancement is larger, with reported enhancement of $10\%$ above the non-rotating Nusselt number \citep{Horn_Shishkina_2015}.

Direct comparison between Nusselt number scalings for dry and moist convection is complicated by the additional buoyancy source from condensation. We define and measure a moist Nusselt number which quantifies the transport of moist static energy, while studies of rotating Rayleigh-B\'enard convection look at a thermal Nusselt number measuring heat transport. For our setup, with a marginally stable thermal stratification, the thermal Nusselt number is undefined since the domain-averaged conductive flux is zero. However, it is still useful to compare general behavior since both quantities are measures of energy transport by convection.

\begin{figure*}
    \centering    \includegraphics[width=\linewidth]{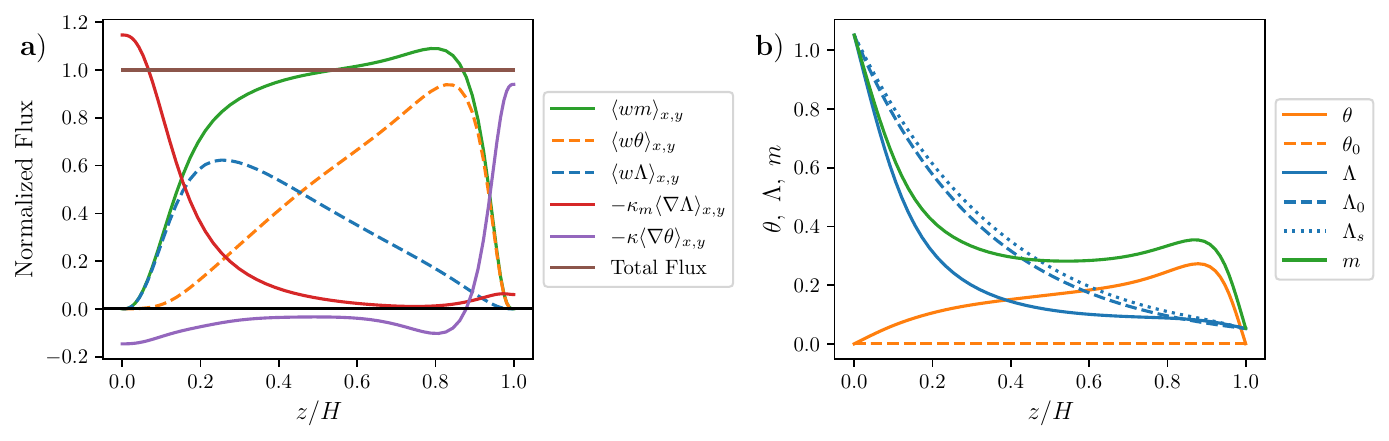}
    \caption{Panel a: Horizontally- and time-averaged vertical profiles of moist static energy fluxes for the simulation shown in Figs. \ref{fig:funnel_cells}--\ref{fig:funnel_vorticity}. The turbulent heat flux (dashed orange line) and the turbulent latent heat flux (dashed green line) sum to form the moist static energy flux (solid blue line). We also show the diffusive thermal flux (purple) and the diffusive moisture flux (red). The total energy flux (brown line) is given by the sum of the solid lines. Panel b: Horizontally- and time-averaged profiles of potential temperature $\theta$, latent heat density $\Lambda$ and moist static energy density $m$. The solid blue line shows the potential temperature of the evolved state and the dashed blue line shows the potential temperature of the initial atmosphere. The solid orange line shows the latent heat of the evolved state, the dotted orange line shows the latent heat at saturation of the evolved state, and the dashed orange line shows the latent heat of the initial atmosphere. Note that the latent heat of the initial atmosphere is identical to the latent heat at saturation and the moist static energy of the initial atmosphere. }
    \label{fig:profiles}
\end{figure*}

To calculate fluxes through the system we add equation \ref{eqn:theta} and equation \ref{eqn:q} to combine the internal energy and latent heat, arriving at an equation for the temporal evolution of the moist static energy $m$,
\begin{equation}
    \frac{D m}{D t} =\kappa\nabla^2  \theta + \kappa_m\nabla^{2}\Lambda \; .
\end{equation}
%%eliminating the condensation term.
This prognostic equation can be written in conservative form,
\begin{equation}
    \frac{\partial m}{\partial t} + \nabla \cdot (m\mathbf{u}  -  \kappa \nabla \theta - \kappa_m \nabla \Lambda) = 0
\end{equation}
where an advective flux of moist static energy, $m\mathbf{u} $, conductive thermal flux, $\kappa \nabla \theta$, and conductive latent heat flux, $ \kappa_m \nabla q$ appear in the divergence term. The advective moist static energy flux can be broken into thermal, $\theta\mathbf{u} $, and latent heat, $ \Lambda\mathbf{u} $, components. In Fig.~\ref{fig:profiles} we plot horizontal averages of the vertical components of these fluxes against depth as well as buoyancy, specific humidity, and moist static energy for a simulation in the funnel regime. Since $m$ is a conserved quantity, the horizontally integrated fluxes of a steady-state solution sum to a constant value, and Fig \ref{fig:profiles} shows how the total flux is partitioned as a function of depth. We find diffusive boundary layers at the top and bottom of the domain. The lower boundary layer is dominated by a large upwards diffusive moisture flux, and the upper boundary layer is dominated by a large diffusive thermal flux. In the middle of the domain advective fluxes dominate, with the advective moist static energy flux growing with height except in the upper boundary layer. In the lower half of the domain this advective flux is largely from the latent heat, but in the upper half of the domain thermal advection dominates as condensation converts moisture into heat.

We measure volume averaged Reynolds and Nusselt numbers for our suite of simulations. The Reynolds number is given by
\begin{equation}
    \mathrm{Re}=\frac{u_{rms}H}{\nu}. 
\end{equation}
where $u_{rms}$ is the root-mean-squared averaged velocity. We find a dynamical Rossby number by taking 
\begin{equation}
    \rm{Ro}=\rm{Re}~\rm{Ek}.
\end{equation}
We define the ``moist'' Nusselt number as the ratio of the total energy flux to the sum of the diffusive fluxes (temperature and moisture diffusion) averaged over the spatial domain. We then use our boundary conditions to simplify the expression for the volume averaged quantity to find 
\begin{equation}
    \mathrm{Nu}_m = 1+\frac{\langle w m\rangle_{x,y,z}}{- \kappa \Delta \theta - \kappa_m\Delta \Lambda}.
\end{equation}
Here angle brackets indicate an spatially averaged quantity.

\begin{figure*}
     \centering
     \includegraphics[width=0.95\textwidth]{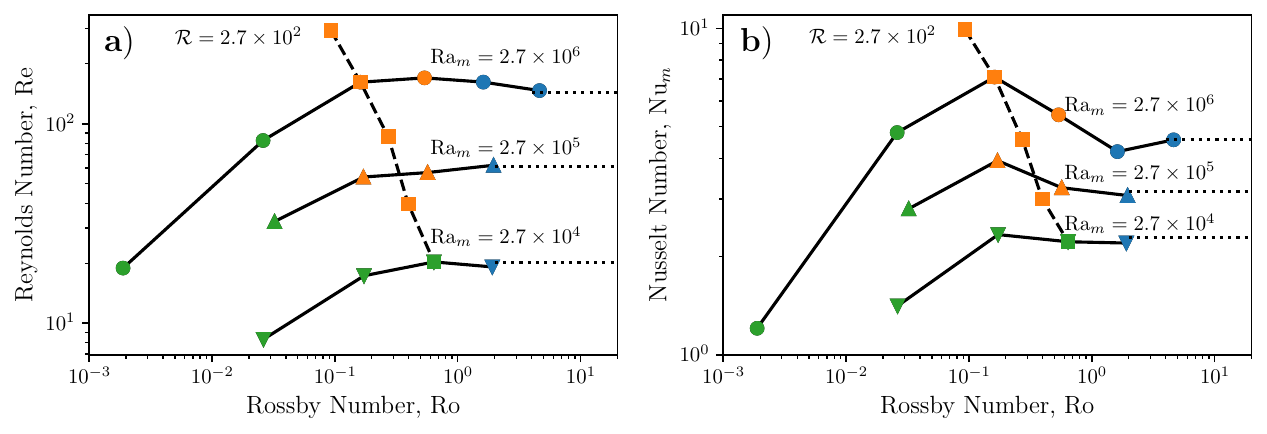}
     \caption{Panel a: volume averaged Reynolds number plotted against Rossby number for simulations at constant values of Rayleigh number (circles and upward/downward triangles) as well as simulations with constant $\rm{Ra}~\rm{Ek}^{4/3}$ (squares and a dashed line). Panel b: volume averaged moist Nusselt number plotted against Rossby number for the same sets of simulations. The dynamical regimes are identified by color, with cells in green, funnels in orange, and plumes in blue. The dotted lines show the values for non-rotating simulations at each of the three constant Rayleigh number cuts.}
            \label{fig:scalings}
\end{figure*}

In Fig. \ref{fig:scalings} we plot scalings of the Reynolds and Nusselt numbers with Rossby number for three paths through parameter space at constant Rayleigh number and for one path at constant reduced Rayleigh number. Along paths of constant Rayleigh number we find that the Reynolds number tends towards zero and the Nusselt number trends towards unity in the limit of fast rotation ($\rm{Ro} \ll 1$). In the slowly rotation limit both the Reynolds and Nusselt numbers approach the non-rotating values, shown by the dotted black lines. However, at intermediate Rossby numbers corresponding to the funnel regime, the Nusselt number is enhanced above the non-rotating values. This enhancement reaches a maximum at approximately $\rm{Ro}=0.1$ in each constant Rayleigh number slice. The magnitude of this enhancement grows with Rayleigh number and reaches $55\%$ in the highest constant Rayleigh number slice, which is substantially larger than the enhancements reported in \citet{Horn_Shishkina_2015}. There is not an equivalent enhancement in Reynolds number, indicating that the increased heat transport does  not arise from faster flows, but instead comes from a more favorable correlation between vertical velocity and the thermal and latent heat densities. 

\section{Discussion} \label{sec:discussion}

%We find three dynamical regimes in our scan of Ek-Ra parameter space. Our cellular and plume regimes are analgous to the cellular and plume regimes observed in (dry) Rayleigh-Bénard convection \citep{Stellmach_etal_2014_rbc_regimes}. However, our funnel cell regime does not have a clear analog. The most comparable regime is Taylor columns, as these fall between cellular and plume regimes and are found at Prandtl number 1. Our funnels consist of an inner descending core sheathed in a hot up-flow. This bears some resemblance to the sheathed flows of Taylor columns, however the inner cores of our funnels only reach to the mid-plane while the upflows span the entire domain. Additionally, our funnels do not drift freely through the domain like Taylor columns, however this may be from our no-slip boundary conditions. 
We identify three morphological regimes of rotating moist convection, a cellular regime, a funnel regime, and a plume regime. Our cellular regime is found at low values of supercriticality, and much like the cellular regime of Rayleigh-B\'enard convection \citep{Stellmach_etal_2014_rbc_regimes, JULIEN_KNOBLOCH_1998, Chandrasekhar_hyro_1961}, we observe vertically coherent, laminar columns in a regular hexagonal packing that spans the entire vertical domain. Unlike Rayleigh-B\'enard convection, however, all of our columns are upflows and share the same thermal fluctuation sign and vorticity due to the asymmetry inherent to the equations as condensation is only consistently triggered in upflows. Furthermore there is a long range corelation between the cells beyond the simple hexagonal packing. Each cell is slightly elliptical, and the orientation of the ellipticity repeats across tessellated clusters of three cells. This is most apparent in panel b of figure \ref{fig:cell_vorticity} where one such cluster of three cells is found in the bottom left corner (approximately at $x=0$ and $y=0.5$) and another with shared orientation found immediately above at approximately $x=0$ and $y=2$. The clusters of three cells fill the whole domain with shared orientation of the cell ellipticities.

\begin{figure}
    \centering
    \includegraphics[width=\columnwidth]{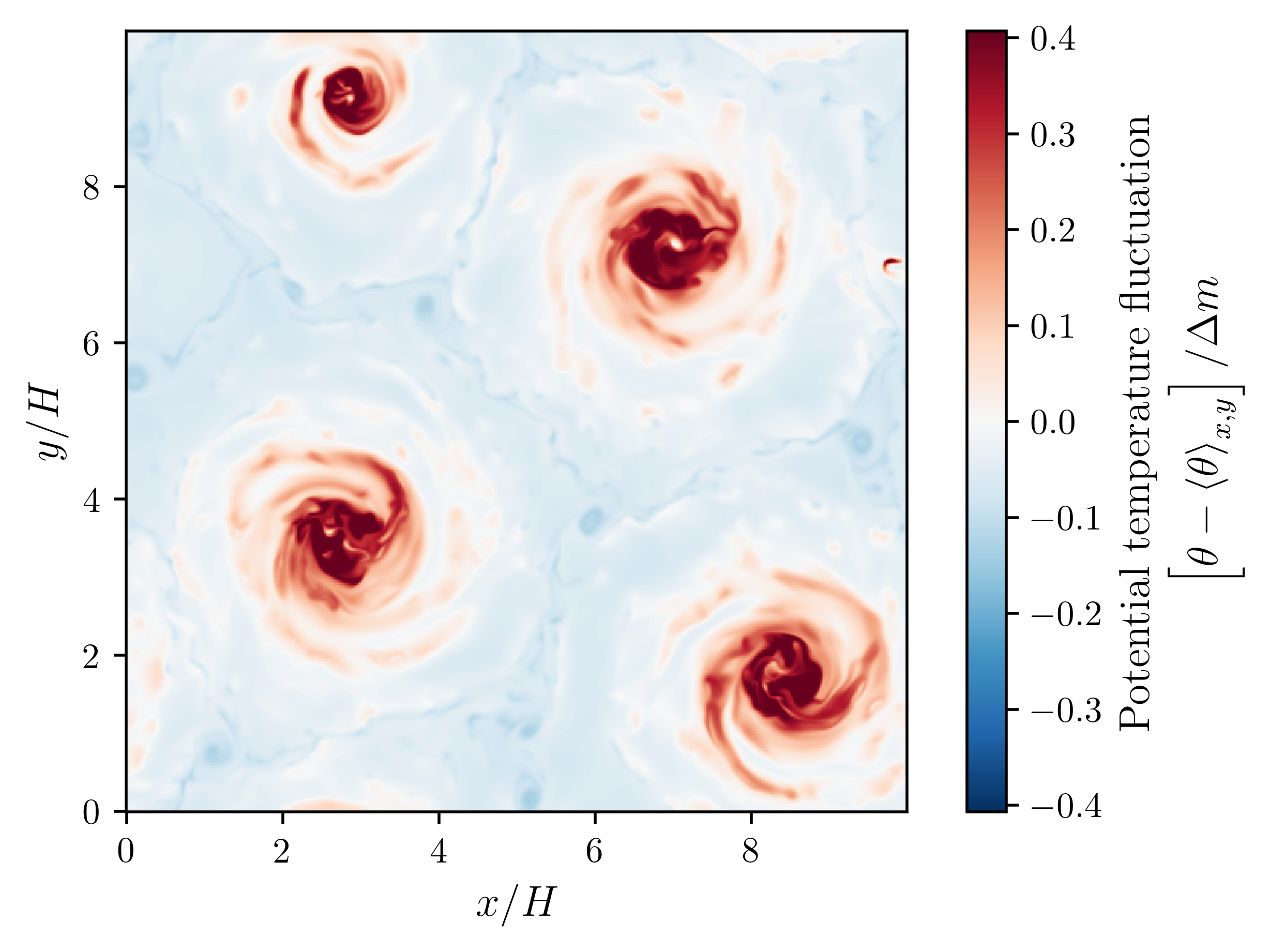}
    \caption{A simulation in the funnel regime with $\mathrm{Ra}_m = 8.49\times 10^6$ and $\mathrm{Ek}=10^{-3}$. The flows form a cellular structure with hexagonal packing, and the funnels now take on a more turbulent structure with spiral arms of hot fluid and an eye of cooler fluid at the center of the hot upflow structure.}
    \label{fig:turbulent_funnels}
\end{figure}

The funnel regime is characterized by upflow columns which span the whole domain with a central core of descending, cold, dry fluid which extends to roughly the midplane. The funnels are packed in a regular hexagonal grid and do not drift through the domain. The regions between the funnels consist of downflow lanes, with vortexes at the vertices of the hexagonal cells. This regime does not have a clear analog in rotating Rayleigh Bénard convection. Convective Taylor columns are found between plume and cell regimes of rotating Rayleigh B\'enard convection, however those structures are vertically symmetric about the midplane and require up-down symmetry \citep{Stellmach_etal_2014_rbc_regimes, Grooms_2010_CTC}. The funnel regime persists at higher supercriticality than the case shown in Figs. \ref{fig:funnel_cells_stream} - \ref{fig:funnel_vorticity}, as can be seen in the more turbulent case in Fig. \ref{fig:turbulent_funnels} where the hexagonally-packed cellular structure persists, but the clearly defined inner funnel vortex structure has transitioned to a more turbulent spiral-arm type structure. The mechanism behind the formation of the funnel structures is not known, however the spiral structures resemble a baroclinic instability. Furthermore the funnel structure may be the result of the breakup of the elliptical, ringed vorticity structure seen in the cellular regime in Fig. \ref{fig:cell_vorticity} at higher Rayleigh number.

The regime of moist convection plumes that we observe is characterized by the breakdown of vertically coherent flows and the lack of a regular geometric structure. This regime appears analogous to the plume regime of rotating Rayleigh Bénard convection \citep{Stellmach_etal_2014_rbc_regimes} with asymmetry between upflows and downflows due to the saturation profile breaking up-down symmetry.

With the suite of simulations used in this study it is difficult to estimate where the bounds of the funnel regime lie in low Ekman, high $\mathcal{R}$ parameter space, and therefore which astrophysical systems are likely to exhibit such dynamics. A regime similar to the geostrophic turbulence regime found in dry, Boussinesq convection studies likely exists, however achieving this regime with direct numerical simulations would be numerically challenging. Dry geostrophic turbulence requires smaller Ekman numbers ($\mathrm{Ek}\leq 10^{-6}$) and sufficiently large supercriticality ($\mathcal{R} =\mathcal{O}(100)$ \citep{JulienGT2012}, and similar values of Ekman and reduced Rayleigh number are likely required for the moist system.

In this study we utilized fully saturated, no-slip boundary conditions, which is a limitation of this study. No-slip boundary conditions were chosen for their simplicity, and for consistence with VPT19, however a no-slip boundary, especially at the upper boundary is highly idealized. Stress-free boundary conditions are more appropriate for gas giant systems, and these boundary conditions may alter the flow morphologies we observe. Additionally, a sub-saturated lower boundary is also more appropriate for these systems. In a system which is stable to dry convection, a subsaturated lower boundary will result in a stably stratified layer in the bottom of the domain, and a convective layer in the upper domain. This will likely produce more complex dynamics such as convective overshoot and the driving of internal waves in the stably stratified layer.  

Estimates place the Rossby number of Jupiter's weather layer at  $\mathrm{Ro}\approx 0.1$ \citep{Guillot_FransTextbook_2004jpsm.book...35G}. This matches the Rossby numbers where we observe the funnel regime. This value of Rossby number does not correspond to a rapidly rotating regime, so while the Rayleigh number of Jupiter is much higher than we can achieve in simulations, the funnel regime may still be relevant. It is notable that this is a regime with unique dynamics with no clear analog in dry convection and with enhanced heat transport properties.

\section{Acknowledgements}
W.T.P. was supported by a NASA FINESST fellowship under grant number 80NSSC22K1850. A.E.F. was supported by an NSF Astronomy and Astrophysics Postdoctoral Fellowship under award no. AST-2402142, with additional support from NSF award no. OCE-2023499 and NASA HTMS grant 80NSSC20K1280. E.H.A.  was supported by NSF grant PHY-2309135 and Simons Foundation grant (216179, LB) to the Kavli Institute for Theoretical Physics (KITP). J.S.O. was supported by DOE EPSCoR grant number DE-SC0024572. Computational resources were provided by NASA HECC on allocations  s2276 and s2867. The authors would also like to thank Bradley Hindman, Ian Grooms, Steve Tobias, Geoff Vallis, and Keith Julien for their helpful conversations on this topic.

%\bibliographystyle{}
%\bibliography{refs}

\newpage

\appendix
\section{Nondimensionalization} \label{sec:ApdxA}
%\section{Appendix A: Nondimensionalization} \label{sec:ApdxA}

To non-dimensionalize the moist fluid equations we scale lengths with the vertical extent of our domain, $H$, such that $\nabla = H^{-1} \widetilde{\nabla}$. We use tildes to denotes a non-dimensional quantity. We choose the thermal diffusion time for our characteristic timescale, $t_D = H^2/\kappa$. This sets $\frac{D}{Dt} = \frac{1}{t_D} \frac{D}{D\widetilde{t}}$ and $\tau=t_D \widetilde{\tau}$. Implicitly this sets our velocity scale, $U = \kappa/H$ and we can write $\mathbf{u} = U \widetilde{\mathbf{u}}$. We choose our pressure scale to be $U^2$, therefore $\phi = U^2 \widetilde{\phi}$. We choose our temperature scale from the change in moist static energy across the domain $\Delta m$, thus $\theta = \Delta m \widetilde{\theta}$, furthermore as $m$ and $\Lambda$ also have units of temperature ${m}=\Delta m \widetilde{m}$ and $\Lambda = \Delta m \widetilde{\Lambda} $. 

We can use the change in moist static energy to write a freefall time, $t_{ff}^2 = H/(g\alpha\Delta m)$ which accounts for the change in buoyancy from the thermal boundary conditions as well as the buoyancy gained from condensation as a fully saturated parcel is displaced from the bottom to the top of the domain. From this we can write a `moist Rayleigh number' 
\begin{equation}
    Ra_m = \frac{g\alpha\Delta m H^3}{\kappa\nu} = \frac{g\alpha(\Delta \theta + \Delta \Lambda)H^3}{\kappa\nu}. 
\end{equation}
Note that this moist Rayleigh number is the sum of the standard, `thermal' Rayleigh number from Rayleigh-Benard convection and a term from moisture. Our moist Rayleigh number also reduces to the standard Rayleigh-Bénard Rayleigh number if $\Delta \Lambda=0$. For our choice of isothermal boundary conditions, the thermal Rayleigh number is zero since the system would be stable without condensation.

In a more general case where $\Delta \theta\neq0$ it is useful to consider the quantity $\Delta \Lambda/\Delta\theta$. If this parameter is zero then there is no latent heat release and we recover the Boussinesq equations, if $\Delta \Lambda/\Delta\theta < 0$ then a convecting system is stable to dry convection but unstable to moist convection, and if $\Delta \Lambda/\Delta\theta < 0$ then the system is unstable both instabilities. If $\Delta \theta=0$, as it is in this study, then the parameter is undefined and the system is marginally stable to dry convection. The magnitude of this parameter determines the relative importance of latent heat to the dynamics much as the parameter $\gamma$ in VPT19 does.

By introducing a Prandtl number $\rm{Pr}={\nu}/{\kappa}$ a Lewis number $\mathrm{Le}={\kappa_m}/{\kappa}$, and Ekman number $\mathrm{Ek} = {\nu}/({2\Omega H^2})$ we can now write the non-dimensional equations:
\begin{equation}
    \frac{D\widetilde{\mathbf{u}}}{D\widetilde{t}} = -\widetilde{\nabla} \widetilde{\phi} + \rm{Ra}_m \widetilde{\theta} \hat{z} + \rm{Pr} \widetilde{\nabla}^2\widetilde{\mathbf{u}} + Ek^{-1}Pr ~ \hat{z} \times{\widetilde{\mathbf{u}}}
\end{equation}
\begin{equation}
    \frac{D\widetilde{\theta}}{D\widetilde{t}} =  \frac{\widetilde{\Lambda}-\widetilde{\Lambda}_s}{\widetilde{\tau}}\mathcal{H}(\widetilde{\Lambda}-\widetilde{\Lambda}_s) + \widetilde{\nabla}^2 \widetilde{\theta}
\end{equation}
\begin{equation}
    \frac{D\widetilde{\Lambda}}{D\widetilde{t}}=-\frac{\widetilde{\Lambda}-\widetilde{\Lambda}_s}{\widetilde{\tau}}\mathcal{H}(\widetilde{\Lambda}-\widetilde{\Lambda}_s) + \mathrm{Le}\widetilde{\nabla}^2\widetilde{\Lambda}
\end{equation}
\begin{equation}
    \nabla\cdot\widetilde{\mathbf{u}}=0.
\end{equation}
Our non-dimensional expressions for temperature and saturation specific humidity are
\begin{equation}
    \widetilde{T} = \widetilde{\theta} - \widetilde{B} \widetilde{z}
\end{equation}
\begin{equation}
    \widetilde{\Lambda}_s=  e^{\widetilde{A} \widetilde{T}}
\end{equation}
where $\widetilde{B} = \frac{g^2 \alpha H}{c_p \Delta m}$ and $\widetilde{A} = \frac{L \Delta m}{R_v \theta_0^2 g \alpha}$. Here $R_v$ is the gas constant for water vapor and $\theta_0$ is the temperature at the base of the domain.  Note that because we nondimensionalize our equations around $\Delta m$ rather than $\Delta T$ as was done in VPT19, $\widetilde{A}$ and $\widetilde{B}$ differ from $\alpha$ and $\beta$ in VPT19. Our simulations use $\widetilde{A} = 0.542$, $\widetilde{B}=5.54$. This is equivalent to $\alpha_\mathrm{VPT} = 3$,  $\beta = 1$, and $\gamma=0.19$ with the nondimensionalization used in VPT19. 

To convert the parmeters used in VPT19, ($\alpha_\mathrm{VPT}$, $\beta$, $\gamma$, $\rm{Ra}_{VPT}$, and $\Delta T$) to those used in this study we first calculate 
\begin{eqnarray}
    \Delta\theta &=& \Delta T + \beta \\
    \Delta\Lambda &=& \frac{\gamma q_0}{g\alpha}[e^{\Delta T \alpha_\mathrm{VPT}}-1]\\
    \Delta m &=& \Delta\theta +\Delta\Lambda.
\end{eqnarray}
Here we assume that the domain in $z$ extends from 0 to 1. Now we can express our nondimensional numbers as
\begin{eqnarray}
    \widetilde{A} &=& \alpha_{\rm VPT} \frac{\Delta m}{\Delta T} \\
    \widetilde{B} &=& \beta \frac{\Delta T}{\Delta m}\\
    \mathrm{Ra}_\mathrm{VPT} &=& \mathrm{Ra}_m \frac{\Delta m}{\Delta T}.
\end{eqnarray}
and $\Delta \Lambda/\Delta\theta$ as simply the ratio calculated from the parameters in VPT19. Note that $\widetilde{A}\widetilde{B}=\alpha_\mathrm{VPT} \beta$ which ensures that the saturation curve does not change with the change in parameters, for fixed thermal boundary conditions.

\newpage
\section{Table of Simulations} \label{sec:ApdxB}
This table lists moist Rayleigh number ($\mathrm{Ra}_m$), Ekman number ($\mathrm{Ek}$), reduced Rayleigh number ($\mathcal{R}$), resolution ($n_x\times n_y \times n_z$), Reynolds number ($\mathrm{Re}$), moist Nusselt number ($\mathrm{Nu}_m$), and morphological regime for every simulation used in this study.
\begin{table}[h]
    %\centering
    \begin{tabular}{c|c|c|c|c|c|c}
        $\mathrm{Ra}_m$ & $\mathrm{Ek}$ & $\mathcal{R}$ & Resolution & $\mathrm{Re}$ & $\mathrm{Nu}_m$ & Regime \\
        \hline
        $2.71\times10^4$ & $\infty$ & $\infty$ & $128\times128\times32$ & 20.06 & 2.29 & Non-Rotating \\
        $2.71\times10^4$ & $1.00\times10^{-1}$ & $1.26\times 10^3$ & $128\times128\times32$ & 19.15 & 2.20 & Plumes \\
        $2.71\times10^4$ & $3.16 \times10^{-2}$ & $2.71\times 10^2$ & $128\times128\times32$ & 20.29 & 2.22 & Cells \\
        $2.71\times10^4$ & $ 1.00\times10^{-2}$ & $5.83\times 10^1$ & $128\times128\times32$& 17.29 & 2.34 & Cells \\
        $2.71\times10^4$ & $3.16 \times10^{-3}$ & $1.26\times 10^1$ & $128\times128\times32$ & 8.30 & 1.41 & Cells \\
        $1.26\times10^5$ & $1.00\times10^{-2}$ & $2.71 \times 10^2$ & $256\times256\times64$ & 39.60 & 3.00 & Funnels \\
        $2.71 \times 10^5$ & $\infty$ & $\infty$ & $256\times256\times64$ & 60.95 & 3.17 & Non-Rotating \\
        $2.71 \times 10^5$ & $3.16\times10^{-2}$ & $2.71\times10^3$ & $256\times256\times64$  & 61.96 &3.08 & Plumes \\
        $2.71 \times 10^5$ & $1.00\times10^{-2}$ & $5.83\times10^2$ &$256\times256\times64$  & 56.96 & 3.25 & Funnels \\
        $2.71 \times 10^5$ & $3.16\times10^{-3}$ &$1.26\times10^2$ & $256\times256\times64$  & 54.08 &3.94 & Funnels \\
        $2.71 \times 10^5$ & $1.00\times10^{-3}$ & $2.71 \times 10^1$ & $256\times256\times64$  & 32.33 & 2.80 & Cells \\
        $5.83 \times 10^5$ & $3.16 \times 10^{-3}$ & $2.71 \times 10^2$ &$256\times256\times64$  & 86.10 &4.57 & Funnels\\
        $8.49 \times 10^5$ & $1.00 \times 10^{-2}$ & $8.49 \times 10^1$ &$384\times384\times96$& 76.07 &4.98 & Cells \\
        $2.71 \times 10^6$ & $\infty$ & $\infty$ & $384\times384\times96$ & 144.28 & 4.57 & Non-Rotating \\
        $2.71 \times 10^6$ & $3.16 \times 10^{-2}$ & $2.71 \times 10^4$ & $384\times384\times96$ & 146.78 &4.56 & Plumes \\
        $2.71 \times 10^6$ & $1.00 \times 10^{-2}$ &$5.83\times10^3$ & $384\times384\times96$ & 161.84 &4.19 & Plumes \\
        $2.71 \times 10^6$ & $3.16\times10^{-3}$ & $1.26\times10^3$ &$384\times384\times96$ & 169.96 &5.44 & Funnels \\
        $2.71 \times 10^6$ & $1.00\times 10^{-3}$ & $2.71\times 10^2$ &$384\times384\times96$ &161.68 &7.10 & Funnels \\
        $2.71 \times 10^6$ & $3.16 \times 10^{-4}$ & $5.83\times10^1$ & $384\times384\times96$ &82.46 &4.80 & Cells \\
        $2.71 \times 10^6$ & $1.00 \times 10^{-4}$ & $1.26 \times 10^1$ & $384\times384\times96$ & 18.90 &1.21 & Cells \\
        $8.49 \times 10^6$ & $1.00 \times 10^{-3}$ & $8.49\times 10^2$ & $512\times512\times128$& 283.41 & 8.61 & Funnels\\
        $1.26 \times 10^7$ & $3.16 \times 10^{-4}$ & $2.71 \times 10^2$ & $512\times512\times128$ & 293.94 & 9.93 & Funnels \\
        $2.71 \times 10^7$ & $1.00 \times 10^{-4}$ & $1.26 \times 10^2$ & $512\times512\times128$ & 234.04 & 8.04 & Cells

    \end{tabular}
    \label{tab:sims}
\end{table}

\end{document}